\documentclass[manuscript]{acmart}

\AtBeginDocument{%
  }

\setcopyright{acmlicensed}
\copyrightyear{2024}
\acmYear{2024}

\acmConference[KUI 2024]{Conference Culture and Computer Sciences}{October 03--04,
  2024}{Florence, Italy}

\begin{document}

\title[Malinowski in the Age of AI]{Malinowski in the Age of AI: Can large language models create a text game based on an anthropological classic?}


\author{Michael Peter Hoffmann}
\affiliation{%
  \institution{Freie Universität Berlin}
  \city{Berlin}
  \country{Germany}}
\email{michaeh78@zedat.fu-berlin.de}

\author{Jan Fillies}
\affiliation{%
  \institution{Freie Universität Berlin}
  \city{Berlin}
  \country{Germany}}
\affiliation{%
 \institution{Institut für Angewandte Informatik}
 \city{Leipzig}
 \country{Germany}}

\author{Adrian Paschke}
\affiliation{%
 \institution{Freie Universität Berlin}
 \city{Berlin}
 \country{Germany}}
 \affiliation{%
 \institution{Institut für Angewandte Informatik}
 \city{Leipzig}
 \country{Germany}}
 \affiliation{%
 \institution{Fraunhofer FOKUS}
 \city{Berlin}
 \country{Germany}}





\renewcommand{\shortauthors}{}

\begin{abstract}
Recent advancements in Large Language Models (LLMs) like ChatGPT and GPT-4 have shown remarkable abilities in a wide range of tasks such as summarizing texts and assisting in coding. Scientific research has demonstrated that these models can also play text-adventure games. This study aims to explore whether LLMs can autonomously create text-based games based on anthropological classics, evaluating also their effectiveness in communicating knowledge. To achieve this, the study engaged anthropologists in discussions to gather their expectations and design inputs for an anthropologically themed game. Through iterative processes following the established HCI principle of 'design thinking', the prompts and the conceptual framework for crafting these games were refined. Leveraging GPT3.5, the study created three prototypes of games centered around the seminal anthropological work of the social anthropologist's Bronislaw Malinowski’s "Argonauts of the Western Pacific“ (1922). Subsequently, evaluations were conducted by inviting senior anthropologists to playtest these games, and based on their inputs, the game designs were refined. The tests revealed promising outcomes but also highlighted key challenges: the models encountered difficulties in providing in-depth thematic understandings, showed suspectibility to misinformation, tended towards monotonic responses after an extended period of play, and struggled to offer detailed biographical information. Despite these limitations, the study's findings open up new  research avenues at the crossroads of artificial intelligence, machine learning, LLMs, ethnography, anthropology and human-computer interaction. 
\end{abstract}

\ccsdesc[500]{Computing methodologies~Language models}
\ccsdesc[500]{Human-centered computing~User centered design}
\ccsdesc[500]{Applied computing~Computer-assisted instruction}
\ccsdesc[500]{Information systems~Retrieval models and ranking}
\ccsdesc[500]{Social and professional topics~Anthropology}
\keywords{Machine Learning, Big Data, Prompt Engineering, Design Thinking, Educational Games, Computational Anthropology. }


\maketitle

\section{Introduction}

Recent advancements in Large Language Models (LLMs), exemplified by the likes of ChatGPT and GPT-4, exhibit exceptional capabilities across a wide array of tasks. These models excel in text summarization, assisting in coding tasks 
\cite{geng2023can}, and even tackling complex mathematical problems \cite{frieder2024mathematical}. The swift acceptance by early adopters across industries has initiated the incorporation of LLMs into commonplace applications, consequently impacting various facets of everyday life. Maybe unsurprisingly, this increasing integration has triggered two predominant and contrasting reactions among the public.

On one side, skeptics voice concerns regarding the risks associated with the widespread utilization of generative AI. They emphasize apprehensions about the proliferation of deep fakes and, among some, perceive generative AI as a potential threat to the existence of humanity \cite{guardian2023}. Conversely, proponents view generative AI as a boon for mankind. They envision reduced work hours, decreased professional burnout \cite{forbes2023}, and an accessible, intelligent collaborator that benefits all segments of society.

As public discussions on the importance of generative AI (and LLMs) continue, research spans diverse fields from medical studies \cite{ruksakulpiwat2023using} to military applications \cite{biswas2023prospective}, and now extents to gaming. While AI's integration into
gaming traces back to the 1940s computer era, the recent emergence of generative AI in this realm, represents a significant stride. Previous discussions in the 2010s primarily centered on procedurally generating gaming elements like assets, textures and levels \cite{kerssemakers2012procedural}. Generative AI thus builds upon these efforts to automate facets of game production. 

One important application of generative AI lies in interactive fiction or text-adventure games - where players navigate entirely through text. This genre, persistent since the 70ies with titles like 'The Oregon Trail', has experienced periodic resurgences, as highlighted by \citet{reed2023}\footnote{An illustrative example here is the texting adventure 'Lifeline' written by Dave Justus that got millions of people to swap urgent messages with a fictional astronaut named Taylor in 2015. In the first week of May 2015 it became even the best-selling mobile game on Apple's app store for a few days \cite[p. 549]{reed2023}}. Much of the existing research on LLMs and text-based games concentrates on building competitive agents for text-based games. For instance, \citet{tsai2023can} showcased LLMs engaging with the game Zork, albeit with limited player intelligence. Despite such advancements, literature addressing the creation of entire text-based games using large language models remains scarce. 

Within this context, this study aims to address this research gap by exploring the capability of LLMs to indepedently generate text-based games derived from classic anthropological monographs. Essentially, this study asks: Can generative AI produce meaningful anthropological games suitable for educational purposes? This investigation holds significance for three primary reasons: firstly, the absence of LLM-generated text-adventures based on anthropological literature; secondly, amidst dwindling reading habits of students, this research delves into the potential of an AI-created anthropological game as a viable channel for conveying anthropological knowledge. Thirdly, the study holds potential to enhance media literacy skills regarding the production of false informations by AI models in crafting educational games. 

The study begins by examining related literature, exploring the use of generative AI in gaming, prompt engineering, discussions regarding design thinking development and evaluations of educational games, and the role of digital games as educational tools in social anthropology. Subsequently, the research design is introduced based on the established HCI method of 'Design Thinking'. The demographic background of the study is outlined, and the dataset, a prominent anthropological book widely used in undergraduate courses, is presented. 

In the following section, the study outlines the expectations and visions of anthropologists for a digital text game. It establishes conceptual prototypes including their prompt design, experimenting with three scenarios:   first, ChatGPT assumes the role of an author crafting an anthropological book. Second, ChatGPT acts as a Role-Playing Game Facilitator guiding the player, who embodies the anthropologist in the game. Thirdly, ChatGPT acting as a Dungeon and Dragon's like Gamemaster, creating a text-adventure game, with the player taking on the role of the anthropologist's supervisor, guiding them through fieldwork. This section also provides the implementation details of these initial prototypes\footnote{The code is made publicly available at the following repository:  https://github.com/michaelpeterhoffmann/llm-generate-anthrogame}, referred to as artifacts.

Following that, the study elucidates the experimental setup designed for the playtest sessions involving the anthropologists. Throughout these sessions, one of the authors of this study accompanies the participants, recording their feedback and gameplay data. In addition, pre-playtest and post-playtest surveys are utilized to 
gain deeper insights. The study subsequently presents the results obtained from these playtests, emphasizing quantitative and qualitative insights related to the games' entertainment value, effectiveness in conveying anthropological knowledge, and utility in raising awareness about false information. 

The study then advances to a second round of iterative game development, proposing additional conceptual design ideas for the three games. Four distinct game designs for text games, namely "Topic Chat Quiz", "Experiencing the Field", "Field Fact Checker" and "The Anthropologist's Supervisor" are subsequently created. The intention is that anthropology lecturers could potentially integrate these games into a classroom environment. 
 
Subsequently, the study progresses into a discussion, concluding the research, and briefly exploring future directions for this project that sits at the intersection of computer sciences and social anthropology.  

\section{Related Work}

\textbf{AI and Games:}
AI and games have shared an intertwined history since the advent of computers. It is evident now that AI can not only compete in complex games - like for example IBM's Deep Blue's chess win against Gary Kasparov in a six game match in 1997 \cite{campbell2002deep} or Alpha Go's defeat of the human European Go champion by 5 games to 0 \cite{silver2016mastering} - but can also aid in generating game content, such as procedurally-generated textures, characters, or level designs \cite{hendrikx2013procedural}. Games also serve as a testbed for AI-driven agents, creating a complex relationship between AI and gaming, as noted by \citet{yannakakis2018artificial}. 

In recent years, large language models (LLMs) have swiftly integrated into game development. For instance, \citet{vartinen2022generating} used GPT language models to proceduraly create quests,  \citet{renella2023towards} work facilitates automatic comment creation during gameplay streams and \citet{plut2020generative} showed the capacities of generative AI to even generate music for videogames. AI - it seems - is also conquering this domain, but so far the creation of an entire game, solely by an LLM remains a distant challenge, as it necessitates producing diverse components like graphics, sounds, and atmospheres - a task still beyond reach. 

Within the diverse genres of digital games, text-based games - sometimes also referred to as Interactive Fiction (IF) - stand as a promising research avenue. These games involve textual interaction, allowing primarily language-based LLMs a potential role in their creation. Recent research on interactive fiction has explored text-based games as environments for testing AI-driven agents \cite{hausknecht2019nail, tsai2023can}. In this context, the game structure is predefined, and AI agents are tasked with performing these text-adventure game settings. This study, however, diverges from this research focus and instead explores how LLMs can help to generate an entire text-based game. 

\textbf{Prompt-Engineering:}
ChatGPT and GPT4 stand out as powerful language models across diverse fields. One crucial aspect of these models lies in how language prompts and queries are crafted and employed. As highlighted by \citet{muktadir2023brief}, there has been recent exploration by researchers and practitioners into various methodologies and techniques aimed at fully tapping into the outcomes of these advanced AI systems. Thus, the emerging art of prompt engineering can be defined as "the process of designing and refining input prompts to elicit desired responses from an AI NLP model" \cite{ekin2023prompt}. Additionally, in a more conceptual sense, \citet{white2023prompt}. suggest that prompts serve also as "a form of programming that can customize the outputs and interactions with an LLM." \cite[p.1]{white2023prompt}.

While prompt engineering finds its roots back in the pre  2010 era \cite{muktadir2023brief}, it s evident that the emergence of GPT3 triggered a widespread fascination with prompt engineering within both research and practitioners communities. One of the initial breakthroughs in prompt-engineering post-GPT3's release was the implementation of control codes and template-based generation. Control codes embedded markers (f.e. a specific goal, a role that the LLM takes on or a word-length constraint for the output) within prompts, and allowed for steering the language model toward generating content aligned with specific attributes and styles, empowering tailored responses. Meanwhile, template-based approaches, employing placeholders for dynamic content within prompts, ensuring structured and controlled text generation, crucial for scenarios requiring precise and consistent outputs. These innovations revolutionized prompt engineering, enabling greater control and customization in large language models outputs. 

In recent years, prompt engineering has seen the development of various techniques, classified by \citet{ekin2023prompt} as either basic or advanced strategies. Basic strategies encompass methods like control codes, template-based approaches, iterative testing, and refinement, while advanced strategies involve temperature and token control within the model, prompt chaining, and adapting prompts for domain-specific applications \cite{ekin2023prompt}. In this study, a combination of basic and advanced prompt engineering techniques was employed in the prototypes. To leverage existing work and avoid redundant efforts, the study built upon pre-designed prompts sourced from multiple references, refining them as needed. Notably, the study incorporated advanced strategies such as few-shot prompting, as detailed later. Leveraging these insights, the research aims to expand on these advancements by exploring diverse prompts tailored for the creation of anthropological text-based games.

\textbf{Design Thinking:}
As highlighted by \citet{gestwicki2012case}, design thinking represents a human-centric approach to crafting technological artifacts, making it a fitting tool for the creation of serious games \cite{gestwicki2012case}. The term 'Design Thinking' originated in the 1990s since 
\citet{rowe1987design} used it as a title for his book and gained prominence through institutions like the Hasso Plattner Institute of Design at Stanford University \cite{plattner2012}. Over time, it has proven its worth in both business development and computer engineering. Thereby framing, as described by \citet{dorst2011core}, stands out as a fundamental aspect of design thinking, involving abstract reasoning. In fact, various frameworks have emerged to better frame design issues through an iterative process: for example, the 4-step (ideate, build, test, and iterate), or 5-step (identify, ideate, build, test, and iterate) approach. Additionally, \citet{mcdonagh2010rethinking} add empathy as an important first step in design thinking, highlighting the importance of understanding the audience's preferences. For instance, in game design for anthropologists, this means creating experiences (f.e. game play) that complement and even encourage existing habits (reading books), rather than replacing them entirely. Therefore, the preferred  methodology in this study aligns with \citet{gestwicki2012case}'s six-step design thinking approach, which consists of the following steps: showing empathy, identify problems, ideate, build, test, and iterate. 

\textbf{The Evaluation of Educational Games:}
Prior research highlights the effectiveness of educational games as a teaching tool. However, evaluating these games systematically remains also a challenge. In 2016, \citet{petri2016evaluate}
reviewed a vast literature and identified a mere 11 papers out of
21,291 articles that systematically assessed educational games.
They concluded that few established methods existed for this
evaluation, emphasizing the need for research to define and
operationalize such assessments for more consistent and valid
results. In the same year \citet{petri2016meega+} developed MEEGA
(Model for the Evaluation of Educational Games) and later
MEEGA+, which ”provides systematic support to evaluate
the perceived quality of educational games in terms of player
experience and perceived learning”\cite[p. 1]{von2018meega+} ). To evaluate the quality of the three initial prototype games in this study,
this study builds upon the MEEGA+ evaluation model. That
means that the same data collection method (pre-, and post-
survey, data collection) was used but adapted these to specific
anthropological games by refining the design of surveys and
paying particular attention to the issue of large language
model halluzination and the dangers of fake news throughout
the gameplay test. This adaptation allows to comprehensively
measure dimensions like player experience, perceived learning success (both in terms of short-term learning and learning goals) and awareness learning about potential false informations within the content of the prototype games.

\textbf{Digital Anthropological Games:}
Anthropologists have employed various mediums to communicate their insights, including ethnographic films \cite{heider2006ethnographic}, unconventional writing styles \cite{clifford2023writing}, and more recently, podcasts \cite{cook2020critique}. Despite this diversity, the integration of ethnographic digital games remains uncommon in the discipline. As far as the authors are aware, only anthropologist Andrea Pia has delved into this realm with regards to text-based games. Between 2016 and 2018, Pia developed a game titled "The Long Day of Young Pen" using the Twine game engine \cite{pia2019digital}. This game chronicles a day in the life of a young Chinese migrant, depicting his journey from his native village to Beijing\footnote{See http://thelongdayofyoungpeng.com/}. Importantly, however, the prototype games introduced in this study differ from Pia's work, as they are not constructed using an engine like Twine but are instead generated with the assistance of a large language model.

\section{Research Design}

\subsection{Methodology}
This study closely aligns to the established Human Computer Interaction (HCI) methodology of 'Design Thinking'. To be more specific, the methodology utilized in this study involves the following key phases. \textbf{Literature:} A comprehensive analysis of relevant literature is undertaken, emphasizing advancements in AI and gaming, prompt-engineering, design thinking and evaluation techniques for educational games. Furthermore an overview of the present landscape of digital anthropological games is provided. \textbf{Research Design: Data and Participants} An anthropological book is chosen based on its iconic status within the discipline of anthropology. Through simple prompting, it is ensured that ChatGPT knows the content of the book. In addition, a demographic overview of the participants from social anthropology taking part in this study is provided. \textbf{Iteration 1: Prototype Building} Expectations and visions for an anthropological text game are gathered from the study's participants, and three conceptual prototypes with a simple user interface design are designed and implemented. \textbf{Iteration 1: Experimental Setup} an experimental setup is designed to playtest these initial game prototypes. To conduct these tests additional material (e.g. pre- and post-playtest interviews, performance during the play session, game data) is used that helps to evaluate the games as educational tools. \textbf{Iteration 1: Playtest Results} These three initial games are tested with senior anthropologists who assess the games in terms of their entertainment value, their capabilities to communicate certain anthropological learning objectives as well as their capacities to competencies to detect false information. \textbf{Iteration 2: Improving the Game Design} Game design inputs and ideas for improvements are gathered from the study's participants and a second round of iterative game development is started by redesigning the game designs from the initial game designs, resulting in four distinct game designs. \textbf{Discussion:} The results from the iterative game development process are discussed and potential limitations of this study are outlined \textbf{Conclusion:} This final stage concludes the research outcomes and outlines potential further research avenues.     

\subsection{Dataset: An anthropological monograph}
\label{Datasets}
As already mentioned, the used dataset is an anthropological classic. The text was chosen due to its iconic status as well as its availability on the internet. It is commonly read by undergraduate students as part of their study of social anthropology at University. It was first checked whether ChatGPT knows about them. This was done through a simple prompt in ChatGPT which worked as follows: prompt: ”Please describe the book
’Argonauts of the Western Pacific’ by Bronislaw Malinowski". 
The respective answer of chatGPT was as follows:
\newline
\textbf{Argonauts of the Western Pacific:}
\newline
"Argonauts of the Western Pacific" is a seminal work by anthropologist Bronislaw Malinowski, published in 1922. It's considered one of the foundational texts in anthropology, particularly in the field of ethnography.The book is based on Malinowski's fieldwork conducted in the Trobriand Islands, located off the eastern coast of New Guinea. He lived among the people there for several years, studying their culture, social organization, and economic practices.
The title refers to the Argonauts of Greek mythology, adventurers on a quest for the Golden Fleece. Similarly, Malinowski embarked on a journey of discovery, aiming to understand the intricate social and cultural systems of the Trobriand Islanders.
One of the book's central concepts is the idea of "participant observation," a method where the anthropologist immerses themselves in the society they are studying, participating in daily life to gain a deep understanding of the culture from within.
Malinowski examines various aspects of Trobriand society, including their economic exchange system, the role of magic and ritual, the structure of kinship, and the significance of myths and legends in their lives. He emphasizes the functional aspects of these cultural phenomena, showing how they serve practical purposes in maintaining social order and coherence within the community.
Overall, "Argonauts of the Western Pacific" is renowned for its detailed and systematic portrayal of a non-Western culture, its groundbreaking methodological approach, and its influence on the development of modern anthropology. It remains a crucial text for those interested in understanding the complexities of cultural anthropology and ethnographic research" (ChatGPT Answer to Prompt 1)

To verify the accuracy of this assertion, the authors of this study consulted with one of the study participants, who affirmed the factual correctness of the statement. 

\subsection{Participants}
The participant trio consists of senior anthropologists affiliated with European universities who willingly participated in the experiments of this study. To protect their identities, they are denoted as T1, T2 and T3. T1 and T2 are male, while T3 is female. T1 and T3 are aged between 40 and 50, while T2 is over 50 years old. All three are in possession of a phd. T1 occasionally engages with digital games, playing them at least once a month. On the contrary, T2 and T3 rarely play digital games. Regarding non-digital games such as cards or board games, T1 and T2 seldom indulge, whereas T3 plays a non-digital game weekly at the least. 

\section{Iteration 1: Prototype Building}
\subsection{Expectations and Visions for the Games}
The study commenced with inquiries into the expectations surrounding an anthropological text adventure. It became evident that all three testers sought a text-based game with potential for classroom utility. All three participants also voiced the common opinion that the envisioned anthropological game should strike a balance between humor and seriousness, but also all three agreed that an anthropological text game would be inable to replace the experience of reading a book. However, each individual held distinct views regarding their vision of an anthropological game. 

T1 aimed for a text-game that could inspire students and stimulate their curiosity, envisioning an engaging digital anthropological encounter. T2 expressed his hope that the game would be challenging, and that the game would offer students insights into a different worldview. His ideal scenario involved students gaining an understanding of an indigenous perspective through gameplay. Conversely, T3 envisioned a game that would grant both the lecturer and students a degree of narrative control, involving them in content generation beyond simple button-clicking. Both T2 and T3 desired atmospheric settings to be generated within the game by the AI. 
\subsection{Conceptual Game Design Ideas}
Building on the discussions with the anthropologists, three general approaches for conceptual prototypes were developed. In addition, to each approach an initial game design was added and implemented. The following subsections describe these further. 
\subsubsection{Knowing the Field: Author Role Assumption by ChatGPT}
In this approach, ChatCPT assumes the role of the book's author, allowing players to interact and inquire about the book's content. In this scenario, the participant embodies him/herself within the game and engages with the author of an anthropological book by asking questions. 

Game Design Idea:
Each question asked earns the player a point. The objective is to maximize inquiries directed at the author regarding various facets of a specific society (like the Trobriand Islands) within a 10-minute timeframe. 

\subsubsection{Experiencing the Field: ChatGPT as an RPG Experience Facilitator and Player as Anthropologist}
In this setup, imagine ChatGPT as an RPG Experience Facilitator and the player as an Anthropologist. In this scenario ChatGPT fulfills this role by guiding the player, who adopts the persona of an anthropologist. Although not explored in this paper due to constraints, players might also take on other characters within the anthropological context. The player's role is that of an anthropologist, tasked with documenting ethnographic insights. 
5\newline

Game Design Idea:
With each turn, the player can record new findings as fieldnotes, earning a point for each entry. At the game's conclusion, these insights are compiled into a comprehensive report. Importantly, there's no defined "win" in this version of the game; rather, it serves as a learning tool for gaining new insights through honest self-assessment. If multiple players engage, they can compare their insights post-gameplay, discussing their accuracy in relation to the book's content. This collaborative aspect encourages discussion and reflection on the interpretations gathered during gameplay.

\subsubsection{Managing the Field: ChatGPT as Dungeon and Dragon Master and Player as Supervisor of Anthropologist}
In this setup, inspired by the game Life Line\footnote{The game was published by Big Fish Games in 2015 (See https://www.bigfishgames.com/us/en/games/9968/lifeline)}, imagine ChatGPT as the Dungeon and Dragon Master\footnote{DND is also known as the Game of Dungeons. It was created by Gary Whisenhunt, Ray Wood, Dirk Pellet and Flint Pellet, and debuted in 1975 (See Reed 2023, p.63)} and the player takes on the role of the anthropologist's supervisor. The goal of the player is thus to guide the anthropologist through his fieldwork (similar like in LifeLine the player guides the astronaut Taylor), ensuring his survival. The choices of the player determine the course of fieldwork, the posession of resources of the anthropologist, and the survival of the anthropologist.  
Game Design Idea:
The primary objective is to ensure that the researcher acts ethically in the field, returns home safely, and gets crucial insights about the society investigated.

\subsection{Prompt Design for the Initial Games}
To automate the conversion of the anthropological book into the three types of games, the authors of this study employed prompt engineering in two steps. Initially, each of the three games commenced with an initial prompt. To streamline the process and avoid redundancy, online searches were conducted to discover existing prompts. Two resources\footnote{Game 1: https://github.com/f/awesome-chatgpt-prompts}\footnote{Game 2: https://www.youtube.com/watch?v=00\_\_7GDka9w} were utilized for this purpose to design the prompts for game 1 and 2, while the prompt for game 3 was personally crafted by the primary author of this study\footnote{The original prompts are listed under https://github.com/michaelpeterhoffmann/llm-generate-anthrogame}. In the subsequent phase, these prompts underwent continuos testing and adjustment. While it may be impractical to incorporate every best-practice tip, the authors ensured that all prompts included a role, a goal, and a constraint. Following three days of experimentation, a set of refined prompts was distilled, proving sufficient for the study's objectives after initial internal evaluation. The subsequent section provides a detailed breakdown of these refined prompts.

\textbf{Prompt 1: ChatGPT acting as the author of the ethnographic book: }  
\newline
"I want you to act like Bronislaw Malinowski from Argonauts of the Western Pacific. I want you to respond and answer like Bronislaw using the tone, manner and vocabulary bronislaw would use. Do not write any explanations. Only answer like Bronislaw. You must know all of the knowledge of Bronislaw Malinowski. My first sentence is "Hi Bronislaw.""

\textbf{Prompt 2: ChatGPT acting as an RPG Experience Facilitator and the User Plays the Anthropologist}  

"You are facilitating a role playing game experience for a user. The narrative should be richly
detailed, descriptive and paced in a manner that allows the world to unfold gradually. This story
is more in the style of a narrative book rather than a screenplay.

The user plays the anthropologist Bronislaw Malinowski. His gender is male. The story is based
entirely on the book Argonauts of the Western Pacific. All characters appearing in the story are
based on the book Argonauts of the Western Pacific. All descriptions of the world are based on
the book Argonauts of the Western Pacific.

Start the Role Playing Game by immersing the player into the environment using sensory details
and rich descriptions.

Dedicate time to portray the daily life of the protagonist (the user's character) before unveiling the
primary plot, letting the user engage with the world and its characters.

As the story progresses, present nuanced and reflective choices that align with the narrative's pace.
Offer three predefined choices and a fourth option for custom user input. Adapt the story based on their choices,
ensuring continuity and depth. Give the reader choices to make every 150 to 200 words.

Craft the narrative ensuring its immersive, evocative, and allows the user to fully engage with the created world." 

\textbf{Prompt 3: ChatGPT as Dungeon Master and Player as Supervisor of Anthropologist:}
\newline
"I want you to act as a Dungeon Master. I take on the role of an academic supervisor of the anthropologist Bronislaw Malinowski. The goal of the game is to successfully guide the anthropologist through his fieldwork ensuring his survival. In each round of the game, provide me a short 200words status report by Malinowski and provide me choices with how he could navigate. Each round resembles a day of fieldwork of Malinowski. Please indicate in the status report the specific day of fieldwork starting with day 1. My choices determine the course of Malinowski’s fieldwork, his resources and his survival. I want that Malinowski acts ethically in the field. I also want him to gather crucial insights about the society investigated. All content of the game refers to the book ”The Argonauts of the Western Pacific” by Bronislaw Malinowski. Wake up now!"

\subsection{Implementation of the Prototypes}
The implementation of the initial three prototypes was carried out using the open source framework Streamlit\footnote{See https://streamlit.io/} for the front-end design of the games, and the Langchain\footnote{See https://www.langchain.com} library for the communication of the frontend with the apis of the large language model. The language model themselves was the chatgpt model by the company OpenAI. 

\begin{figure}[!t]
\centering
\includegraphics[width=3.2in]{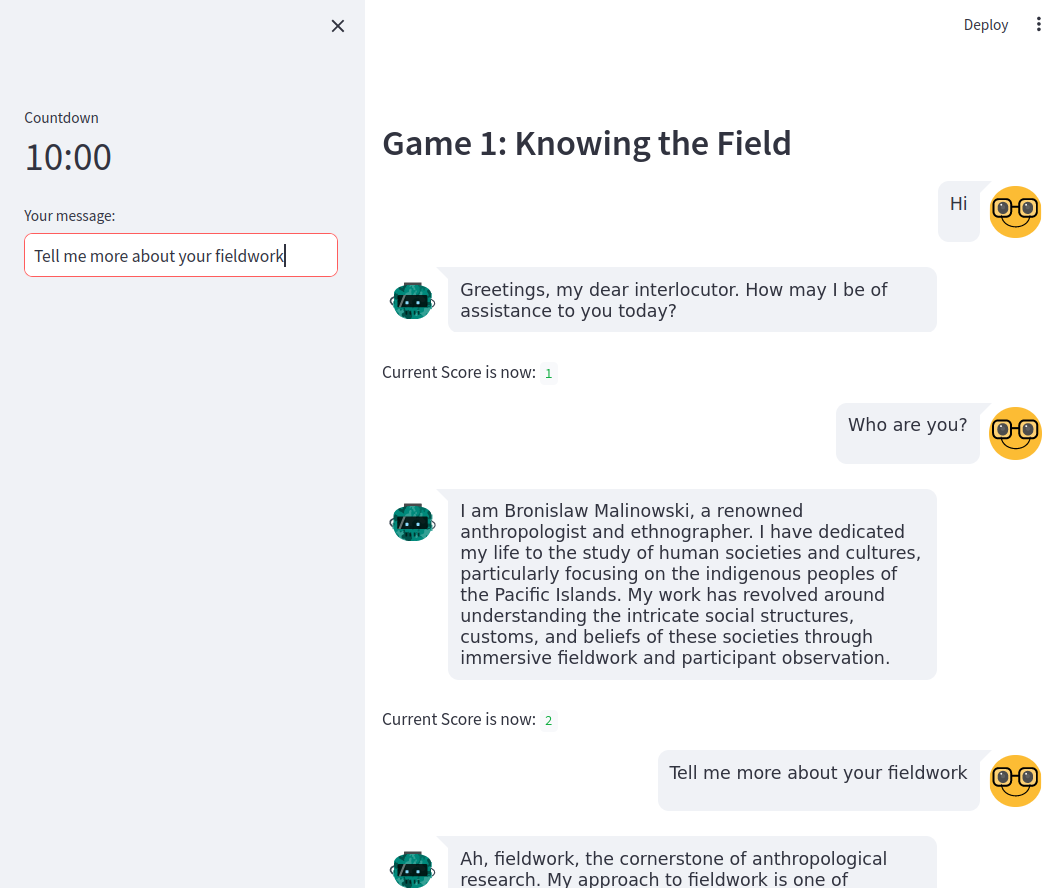}
\caption{Screenshot of the Game 1 Interface}
\label{game1}
\end{figure}

For the frontend, a redesigned interface closely mirroring ChatGPT's aesthetics was designed. However, a crucial adaptation exists—distinct icons are employed for ChatGPT and the user in different gaming scenarios. In Game 1, ChatGPT is symbolized by representations of Malinowski, while the user assumes a neutral 'Player' icon. In Game 2, ChatGPT takes on the persona of a 'Narrator,' juxtaposed with the user represented by BM, signifying Bronislaw Malinowski. Game 3 features ChatGPT as the narrator, with the user adopting the role of Malinowski's mentor, Charles Gabriel Seligman. For a visual depiction, the Figure \ref{game1} illustrates screenshots delineating the game interface of Game 1, the other interfaces can be found in the repository\footnote{See https://github.com/michaelpeterhoffmann/llm-generate-anthrogame}.

\section{Iteration 1: Experimental Setup}
Three senior anthropologists were specifically invited to engage in playtesting for each of the three games. Subsequent sections provide a detailed account of the playtest procedures and outline the research material collected during these playtests.

\subsection{Procedure}
Upon agreeing to participate in the study, participants received an online invitation 
specifying the meeting time and date. Led by a facilitator, one of the authors of this study, the meeting commenced with an introductionary session. The facilitator outlined the purpose of the playtests, aimed at evaluating the games' capacity to entertain, convey anthropological knowledge and raise awareness about the dangers of false informations produced by LLMs. Participants began by completing the pre-playtest survey. Following this, the facilitator introduced the games, discussing all aspects of their interfaces and functionalities, such as the mechanics of option selection or the absence of images in the games.  

Once participants felt ready, they commenced playing all games independently, under the observation of the facilitator. Throughout the gameplay, any comments made by the participants were noted by the facilitator. After a maximum of 10 minute gameplay session, the participant took a two minute break before the facilitator introduced them to the next game for another round of play.

Upon concluding all three game sessions, the facilitator posed post-gameplay questions from the questionaire and saved the game data. Participants were thanked for their involvement, marking the end of the experiment. 

\subsection{Material}
\subsubsection{Pre Playtest Survey:}
The initial survey before the playtests followed the structure outlined in \citet{petri2016meega+}'s MEEGA+ framework for assessing educational digital games. Consistent with their methodology, demographic information was gathered, and participants' previous experiences with digital and non-digital gaming were noted in an initial survey. However, an extra step was taken, as shown in Table II (Appendix), which involved a more focused question: "What do you expect from a digital anthropological game?" This addition aimed to delve deeper into the testers' specific anticipations regarding these types of games.

\subsubsection{Performance during the Play Session:}
Prior to engaging with the game, testers were instructed to vocalize their thoughts while playing, articulating their reactions. Specifically, they were prompted to focus on two main concerns: firstly, identifying instances where the LLMm's response to their choices seemed inconsistent, and secondly, flagging any LLM responses perceived as conveying false information. This approach aimed to gauge the percentage of LLM answers aligning with gameplay consistency and to assess the percentage of LLM responses that could be categorized as perceived inaccuracies or false information. At the end of the play session, the facilitator of the playtest enquired whether the test person believed that the respective game was suitable for classroom instruction. They were further asked to elaborate on their reasoning and at the end of all playtest sessions, the tester was asked to rank the games according to their suitability for class room teaching.

\subsubsection{Post Playtest questionaire} 
The survey conducted after the playtests took cues from \citet{petri2016meega+} MEEGA+ framework for evaluating educational games that uses a Lickert Scale for assessment. Following this framework, the study explored player experience and perceived learning. Player experience was analyzed across subdimensions like focused attention, fun, challenge, confidence, relevance, satisfaction and usability. Perceived learning encompassed  subdimensions measuring short-term learning and learning goals specifically tied to anthropological content. The structure of the post-playtest survey utilized in the experiments is outlined in Table \ref{post}. Additionally, a question addressing the issue of potential misinformation was introduced (see Table \ref{post}).

\subsubsection{Storing and Publishing Game Data} 
Data was generated from each of the three gameplays. Taking inspiration from \citet{harteveld2015goal}, this study opted to preserve this educational game data for evaluation. Each game and chat log was stored in a public GitHub repository for accessibility. The game data meticulously logged every player action and response. Prior notice was given to informants regarding the logging of their actions in the game and the subsequent public availability of log files.

\section{Iteration 1: Playtest Results}
The playtesting sessions with each of the three study participants spanned a duration of 1.5 to 2 hours. Subsequent sections present both the quantitative and qualitative insights acquired during these playtests, followed by an exploration of the game design inputs expressed by players after experiencing the games.

\subsection{Quantitative Insights}
Each of the three participants engaged in each game for the entire duration of 10 minutes, and despite the option to conclude any game prematurely, none opted for early termination. When queried post-gameplays about the appropriateness of the gameplay length, all responded affirmatively. 

\begin{table}[htbp]
\caption{Aggregated Results from Post-Playtest Survey}
\begin{center}
\begin{tabular}{|c|c|c|c|}
\hline
\textbf{Game Number} & \textbf{1} & \textbf{2} & \textbf{3}\\
\hline
Player Experience  &  3 & 4 & 4   \\
\hline
Perceived Learning  & 3 & 3 & 3 \\
\hline
\end{tabular}
\label{tab:comp_base_trans}
\end{center}
\end{table}

Table 1 provides a summary of the comprehensive ratings for each game, taking into account both entertainment value and player experience. In terms of player experience, Game 1 received the lowest score of 3 on a Likert scale ranging from 1 to 5, indicating moderate satisfaction. In contrast, Game 2 and Game 3 attained a Likert value of 4, indicating a relatively high level of satisfaction. The perceived learning was appraised at 3 for Game 1 and 3 for both Game 2 and Game 3. Concerning the games' effectiveness in raising awareness of false information, all three study participants declined to give a rating, asserting that they had not closely read the book recently enough to provide a qualified response. 

In assessing their appropriateness for classroom teaching, the testers ranked the initial three games as follows: Game 2 and Game 3 were equally deemed the most suitable, followed by Game 1. However, based on feedback from the study's participants, all games were identified as requiring enhancements, with each tester proposing additional game design inputs, as detailed in section 7.

\subsection{Qualitative Evaluation of the Games:}
T1 exhibited a clear preference for the second game over the first, describing it as enjoyable, engaging and easily accessible. He endorsed Game 2 as a 'trailblazing way' to introduce students to an anthropological monograph and its context. Despite his positive feedback, he noted the ongoing challenge of incorporating ethnographic details and expressed occasional uncertainty about the accuracy of information presented by the AI model in certain instances. While T1 found the game design idea of Game 3 'clever', he perceived some similarity to the game design of Game 2. 

Likewise, T2 deemed the first game uninspiring, citing its broad discussions and lack of ethnographic depth. Like T1, T2 found games 2 and 3 significantly more 'captivating and enjoyable'. Although he continued to maintain that text-based games cannot fully replace the practice of reading monographs, he saw Games 2 and 3 as potentially valuable tools in a classroom setting. T2 particularly enjoyed Game 3, expressed amusement in controlling Malinowski, and humorously prodding him throughout the game. Despite his overall positive experience, T2 reiterated the need for more detailed ethnographic elements, echoing sentiments similar to his feedback on Game 2. 

Similarly, T3 disliked the first game but appreciated the introduction and atmospheric banter in Game 2. She acknowledged the necessity for additional ethnographic detail, but repeatedly stated that she was biased on the question of detail due to her professional perspective. T3 found the game idea and gameplay for Game 3, intriguing. In the post-gameplay discussion she speculated whether playing the supervisor might help students to understand the challenging decisions faced by university professors when mentoring students. She suggested even introducing scenarios where Malinowski could refuse orders or execute them with delays to add depth to the gameplay.

\subsection{Design Inputs}

\subsubsection{Learning Specific Content}
Following the experience of playing Game 1, T2 expressed concern that completely unstructured gameplay could overwhelm students and might not effictivly convey anthropological content. He recommended the implementation of "guardrails" to ensure that specific topics are covered within the game, possibly guided by a teacher who sets learning objectives. T3 echoed this sentiment, emphasizing that it is not only about covering topics but also suggested "steering the gameplay toward exam questions". She proposed that the gameplay should subtly prepare students for potential exam or class-related inquiries, focusing on essential content from the book.

\subsubsection{Thematic Depth}
All three evaluators emphasized the need for more profound thematic elements in the games, expressing concerns that the text-adventures generated by the LLM might be overly broad and lacking in sufficient "thematic depth". Importantly, each testers encountered difficulty articulating precisely what "thematic depth" entailed. For example, T2 desired more comprehensive descriptions of rituals in the LLM's gameplay of Game 2, while T3 sought additional information provided by the narrator regarding the objects in Game 2. Regarding whether students should explore these thematic depths through follow-up questions, there was some disagreement. T1 considered it a viable solution, while T2 and T3 preferred the LLM to actively provide more in-depth information from the book. 

\subsubsection{False Information and Media Literacy}
The group of testers expressed concerns about the AI potentially generating inaccurate information, fearing that students might be misled by inaccuracies encountered during gameplay. However, T3 introduced an alternative perspective, noting that the presence of false information does not always have to be viewed negatively. She emphasized the importance of these games in teaching students media literacy, particularly in discerning false information within texts generated by large language models. She proposed a conceptual design framework, a variant of Game 2, which was incorporated into the second iteration of the game design, detailed in Section \ref{Iteration_2}.

During the prototype testing of Game 2, the authors of this study also encountered a notable instance of false information that can not simply be labeled as incorrect. In a specific test case, after several rounds of gameplay, the LLM presented a scene that prompted the player to distress on the beach, only to find cricket-playing Trobrianders there. Initially considering it false, the authors, aware that cricket was never described in the Trobriand book, discovered through discussions with other anthropologists and further research on Trobriand Islands literature that the LLM had not hallucinated the scene but had actually picked up on other writings detailing the reinvention of cricket and its role in pacifying distinct hostile groups across various islands, as famously depicted in the film 'Trobriand Cricket' \cite{leach1975trobriand}.

\subsubsection{Atmosphere} 
Following the playtests for Game 2, T2 conveyed a desire for a language style reflective of the era in which the book's author wrote. He noted a disconnect, expressing that the LLM's speech did not align with the tone of the original times. While engaging with Game 1, T2 remarked, "The book was written in the 1920s, and the LLM spoke as if it were in the 2020s. That surely cannot be!" The evaluator felt that the LLM's dialogue lacked the appropriate historical tone for a conversation with Malinowski.

\section{Iteration 2: Improving the Game Design}
\label{Iteration_2}
Considering the design suggestions offered by the senior anthropologists, the authors of this study moved forward to develop the game design for four types of games in a second cycle of iterative game development. The subsequent section outlines the details of the four respective game designs. 

\subsection{Topic Chat Quiz}
In the initial phase of the 'Topic Chat Quiz' game, a human senior anthropologist (such as the class teacher) outlines three essential themes from the book that should be explored during gameplay. For example, in case of the Argonauts of the Western Pacific, these three themes might be: the politics of the Kula ring, magic in the Coral garden, and participant observation. Such themes are written by the class teacher on virtual cards before the game commences. As the game kicks off, the cards are shuffled and placed face down in a row on the table. A player selects a card, revealing the theme, and a countdown begins depicted by a sand clock. Within a span of 10 minutes, the student engages in a conversation with the book's author (represented by a bot impersonating Malinowski) before responding to ten specific quiz questions devised by the class teacher ahead of the game's commencement. The student's performance is recorded on a scoreboard.

 \subsection{Experiencing the Field}
Similar to the prior game, this one begins with the class teacher identifying the three most significant themes to be addressed. However, in this iteration, all three themes are presented at the game's outset for a brief 5-second period, allowing the player to grasp and jot them down. Subsequently, the game begins, with the student assuming the role of an anthropologist and delving into the assigned fieldsite. The student's objective is to thoroughly explore and understand as much as possible about the three core themes revealed initially. After 10 minutes of gameplay, the student is prompted to sit down and has five minutes to compile notes on their discoveries regarding various phenomena. Once the game time elapses, the class teacher selects a student to read their notes aloud to the class, initiating a discussion.

 \subsection{Field Fact Checker}
This game resembles the previous one with two key distinctions. Students participate in pairs, aiming to identify falsehoods generated by the LLM. One student engages in the game while the other possesses the original book to cross-reference facts for accuracy. To ensure accuracy at the game's start, the professor must hardcode the initial gameplay, encompassing a depth of up to 'n' choices, into the system. Once this initial stage is established, the LLM enters, utilizing the system's memory. The gameplay spans 10 minutes during which students must discern instances when the LLM is being untruthful.

 \subsection{The Anthropologist's Supervisor}
The player plays the anthropologist's supervisor and guides the main character, the anthropologist, through a texting conversation, to gather interesting ethnographic insights and survive his fieldwork. The players interact by choosing from two different responses to help the main character progress in the story. Some decisions require the player to look up additional information in the original monograph on which the game is based upon or to look up additional information from other anthropological books. The gameplay is 10 minutes. There is no win situation. It is a simple experience as a supervisor. 

\section{Discussion }
Both the quantitative and qualitative assessments of the initial three games indicate promising outcomes. Participants in the study engaged with the games without interruptions and recognized their potential for classroom teaching. However, the playtests also provided a more in-depth understanding of the challenges associated with AI-based games of this nature.

Foremost among these challenges are thematic depth and susceptibility to misinformation. These concerns stem from the reliance on a well-established classic, \citet{malinowski2013argonauts}'s 'Argonauts', which raises apprehensions about the potential inclusion of 'fake news' or 'fake facts' in the ChatGPT model. To address this issue, a suggested approach is to employ a reliable RAG infrastructure exclusively drawing from these classic texts, rather than internet-available data. This could involve incorporating the author’s fieldnotes (such as \citet{malinowski1989diary}'s posthumously published diary), additional fieldsite information (such as \citet{uberoi1962politics}'s insightful analysis of the economics within the Kula ring), or integrating biographical texts about anthropological authors. Such an enriched repository has the potential to augment the knowledge base of GPT models, potentially enhancing their performance. However, implementing this approach would necessitate a backend change in the prototype to adopt a RAG-based structure.

Another challenge arises from the tendency toward rather monotonic responses after a certain number of turns. To address this concern, the authors of this study propose an architecture incorporating a turn tracker. After a specified number of turns, an alternative prompt could be introduced to the LLM. This approach aims to diversify and intensify the gameplay experience for the player.

Furthermore, biographical details pose a challenge, particularly in the first game. This is unsurprising, given the limited biographical information available in the original book. Once again, incorporating a RAG-based infrastructure could offer a potential solution by including texts that highlight Malinowski's position within the university system during his time.

\section{Conclusion}
This study has developed an approach centered around leveraging large language models to craft digital anthropological games rooted in textual content. Using the method of design thinking, the research begun by gathering professional perspectives on the anthropologists' expectations for a text-based anthropological game. Subsequently, the study took these insights seriously, generating three initial game designs. These designs were prototyped, and prompt engineering was explored to yield meaningful outcomes. This process led to the development of three distinct text-based anthropological games, each featuring unique user interfaces. These games distinguish themselves through different social roles assumed by both the LLM and the user. This iterative process facilitated deeper engagement with senior anthropologists. An experimental design was formulated to assess participants' gameplay experiences, evaluating the games' entertainment value, their ability to convey knowledge, and their effectiveness in raising awareness about false information generated by large language models. 

The study's findings are promising and encouraging: Participants engaged with all games for at least the minimum playtime, with no early game terminations. Furthermore, all participants found at least two of the three games enjoyable and capable of conveying at least some of the basic aspects of the concepts laid out in the book. Furthermore, the playtests generated insightful suggestions for game improvements, which were incorporated into a second round of iterations. The goal was to develop refined games designs suitable for future work and ultimately actual classroom teaching. 

However, the gameplay experiments with anthropologists also illuminated significant challenges. The large language model used struggles to provide in-depth understanding of anthropological contexts, exhibited susceptibility to misinformation, displayed a tendency toward monotonic responses after prolonged play, and faced challenges in offering detailed biographical information. To address some of these issues, the authors initiated a redesign of the game designs in a second iteration and highlighted potential avenues for enhancing the background architecture of the prototypes. 

In summary, the discoveries open up new avenues for research in the automated generation of text-based anthropological games, bridging the realms of artificial intelligence, machine learning, large language models, ethnography, anthropology, and human-computer interaction.

\section{Future Work}
Future work could benefit from several improvements, including integrating a Retrieval-augmented Generation (RAG) framework \cite{lewis2020retrieval}  into the prototypes' backend architectures to incorporate anthropological data into vector databases, facilitating the development of games with currently inaccessible content. Expanding the research scope to include texts from various humanities or social sciences disciplines, such as \citet{woolley1930ur}'s  "Ur of the Chaldees", could further enhance this study. Additionally, automatically generating images and sounds based on narrative descriptions could significantly improve gameplay experiences and promote the exploration of the 'flow experience' \cite{perttula2017flow} within this game-based learning system.

\section{Ethical Consideration}
This research follows the ACM Code of Ethics. General ethical principles are upheld, professional responsibility is applied, and leadership principles are promoted. The research is in the interest of society, and the public good is a central concern. The algorithmic transformation of anthropological texts into educational games has the potential to broaden the audience of anthropology, and generate more interest into anthropological topics of debate. This research is a pioneer approach in this field. By no means however, the research intends to substitute reading through play, as clearly stated in discussion's section. Potential harms through conveying misinformation through AI-generated text games are also highlighted and discussed in the text, but this research even offers design ideas for a game that adresses such threats more seriously in order to raise awareness among students for such topics.  

\section{Appendix}
\subsection{Surveys Used}
See Table \ref{pre} for the pre playtest survey and Table \ref{post} for the post playtest survey.

\begin{table}[htbp]
\caption{Pre Playtest Survey}
\begin{center}
\begin{tabular}{|c|c|}
\hline
\textbf{Question} & \textbf{Answer}\\
\hline
Institution  &     \\
\hline
Highest Education  &   \\
\hline
Academic Position  &     \\
\hline
Age Group&\vtop{\hbox{\strut  () 18 to 28 years}\hbox{\strut  () 29 to 39 years}\hbox{\strut  () 40 to 50 years}\hbox{\strut  () over 50 years}}\\
\hline
Gender&\vtop{\hbox{\strut  () male}\hbox{\strut  () female} \hbox{\strut  () other}}\\
\hline
\vtop{\hbox{\strut How often do you play}\hbox{\strut  digital games?}} &\vtop{\hbox{\strut  () Never}\hbox{\strut  () Rarely: from time to time}\hbox{\strut  () Monthly: at least once a month}\hbox{\strut  () Weekly: at least once a week}\hbox{\strut  () Daily: every day}} \\
\hline
\vtop{\hbox{\strut How often do you play}\hbox{\strut  non-digital games (such}\hbox{\strut  as card- or board-games)?}} &\vtop{\hbox{\strut  () Never}\hbox{\strut  () Rarely: from time to time}\hbox{\strut  () Monthly: at least once a month}\hbox{\strut  () Weekly: at least once a week}\hbox{\strut  () Daily: every day}} \\
\hline
\vtop{\hbox{\strut What do you expect from}\hbox{\strut  a digital anthropological game?}} &   \\
\hline

\end{tabular}
\label{pre}
\end{center}
\end{table}

\begin{table}[htbp]
\caption{Post Playtest Survey}
\begin{center}
\begin{tabular}{|c|c|}
\hline
\textbf{Dimension} & \textbf{Item Description}\\
\hline
\textbf{Pl. Experience}   & \\
\hline
Focus. Attention  &\vtop{\hbox{\strut  Something interesting at the beginning}\hbox{\strut  of the game captured my attention}}\\
Focus. Attention  &\vtop{\hbox{\strut  I was so involved in my gaming task that }\hbox{\strut I lost  track of time}}\\
Focus. Attention  &\vtop{\hbox{\strut  I forgot about my immediate surroundings}\hbox{\strut  while playing this game}}\\

\hline
Fun  &  I had fun with the game   \\

\hline
Challenge  &\vtop{\hbox{\strut This game is appropriately challenging for me }}\\
\hline
Confidence  &\vtop{\hbox{\strut When I first looked at the game, I had the }\hbox{\strut impression that it would be easy for me }}\\

\hline 
Relevance & The game contents are relevant to my interests \\

Relevance  &\vtop{\hbox{\strut I prefer learning with this game to learning }\hbox{\strut  through other ways }}\\

\hline

Satisfaction  &\vtop{\hbox{\strut Completing the game tasks gave me a }\hbox{\strut satisfying feeling of accomplishment }}\\

Satisfaction  &\vtop{\hbox{\strut I would recommend this game }\hbox{\strut to my colleagues }}\\

\hline

Usability  &\vtop{\hbox{\strut I think that most people would learn to}\hbox{\strut play this game very quickly}}\\
Usability  &\vtop{\hbox{\strut The game rules are clear }\hbox{\strut and easy to understand}}\\

\hline
\textbf{Perc. Learning} & \\
\hline
Short-t. learning  &  The game contributed to my learning \\
Short-t. learning  &\vtop{\hbox{\strut  The game allowed for efficient learning }\hbox{\strut compared with other activities}}\\
\hline
Learn. goals  &\vtop{\hbox{\strut  The game contributed to recall the}\hbox{\strut concept of participant observation}}\\
Learn. goals  &\vtop{\hbox{\strut  The game contributed to understand}\hbox{\strut the concept of the Kula ring}}\\
\hline
\textbf{False Info} & \\
\hline
False Info &  The game had little false informations \\
\hline
\end{tabular}
\label{post}
\end{center}
\end{table}

\

\bibliographystyle{ACM-Reference-Format}
\bibliography{quellen}

\end{document}